\documentclass[prb,preprint,amsmath,amssymb]{revtex4}
\begin{document}
\author{Kevin Leung}
\affiliation{Sandia National Laboratories, MS 1415, Albuquerque, NM 87185,
{\tt kleung@sandia.gov}}
\date{\today}
\title{Ion-dipole interactions are asymptotically unscreened by water
in dipolar nanopores, yielding patterned ion distributions}

\input epsf

\begin{abstract}
The permeation, rejection, and transport of electrolytes in water-filled 
nanopores are critical to ion current gating and desalinalion processes in 
synthetic porous membranes and the functions of biological ion channels. 
While the effects of confinement pore polarizability, and discrete channel
charge sites have been much studied, the potentially dramatic impact of
dipole-lined synthetic pores on electrolytes has seldom been addressed.
Dipole layers naturally occur on the interior surfaces of certain
nanopores, leading to intrinsic preference for cations or anions. This
preference can be exploited when the membrane surface is functionalized
differently from the pore interior or when there are alternating
dipolar/nondipolar stretches inside a long pore. The dipole-ion
interaction is asymptotically unscreened by water, leading to ionic,
charge segregated, insulating behavior that can block ion transport,
and potentially novel current-voltage ($I-V$) characteristics.
\end{abstract}
 
\maketitle
 
Water-filled nanopores exhibit a remarkable ability to exclude or selectively
transmit ions.\cite{parsegian,sansom,hummer,hummer1,hansen,pore,charge,kcsa}
They have important biological functions,
and are potentially useful for desalination and other applications.
The effects of dielectric mismatch,\cite{parsegian}
nanopore diameter,\cite{sansom} net charge,\cite{charge}
and polarizability\cite{hummer1,hansen,pore1} on ion permeation
and rejection have been much studied.  Surprisingly, while the celebrated
KcsA channel selectivity filter relies on carbonyl group dipoles pointed
radially into its narrow passageway for its function,\cite{kcsa}
dipolar decorations in synthetic nanopores have received less attention.   A
recent force field-based model has however shown that synthetic silica
nanopores contain dipoles on its interior surface and exhibit an intrinsic
preference for cations --- even at the pH of zero charge.\cite{pore}
This work elucidates the dramatic effect of dipolar layers in nanopores on
electrolyte permeation, which may lead to novel current-voltage ($I-V$)
characteristics, and discusses pore geometries where this effect may
be realized.
 
An infinitely long cylindrical pore of radius $R_d$ with radially aligned
surface dipoles of magnitude $d$ and uniform surface density $\eta_d$ exerts
a uniform electrostatic potential $\phi_{\rm d}= -4\pi \eta_d d$ in the
pore interior, {\it independent} of $R_d$ or radial position of ion $R$.  This
intrinsic $\phi_{\rm d} |e| $ is predicted to be $\sim -24$~kcal/mol in
silica pores,\cite{pore} and exceeds -100~kcal/mol inside the KcsA
filter.\cite{kcsa,support} Here $e$ is the electronic charge.  A
uniform dipole layer thus induces no electric field and does not interact
with uncharged H$_2$O molecules.  Hence its interaction with ions inside
the pore is unmediated (unscreened) by the confined water.  This is
consistent with our hitherto puzzling observation that the 
free energy preference for Na$^+$ over Cl$^-$
in the infinitely long silica pore of Ref.~\onlinecite{pore} is 
fairly independent of the presence of water.  Thus the simple
uniform dipole density in our model apparently captures essential
nanopore-ion interactions in nanoporous silica, where H$_2$O molecules do
form hydrogen bonds with surface silanol groups.  This model
should be pertinent to other inorganic nanopores too.
 
\begin{figure}
\centerline{\epsfxsize=3.40in \epsfbox{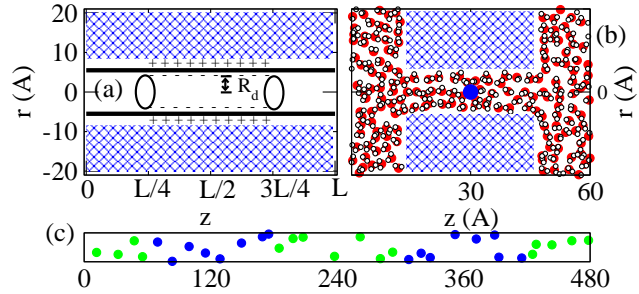}} 
\caption[]
{\label{fig1} \noindent
(a) Cylindyrical nanopore unit cell for the ``tube'' geometry.  
Ions and water are excluded from the crossed regions.  Radial dipole
moments (``+/-'') exist on alternating stretches along the pore
surface.  (b) ``Membrane'' simulation unit cell.  Blue and red/white
spheres depict Na$^+$ and H$_2$O respectively.  Dipoles exist
on interior pore surface only.  
(c) Charge segregated insulating behavior in tube geometry
(see text).  Na$^+$ and Cl$^-$ are depicted as blue and green spheres.}
\end{figure}

\begin{figure}
\centerline{\hbox{\epsfxsize=3.60in \epsfbox{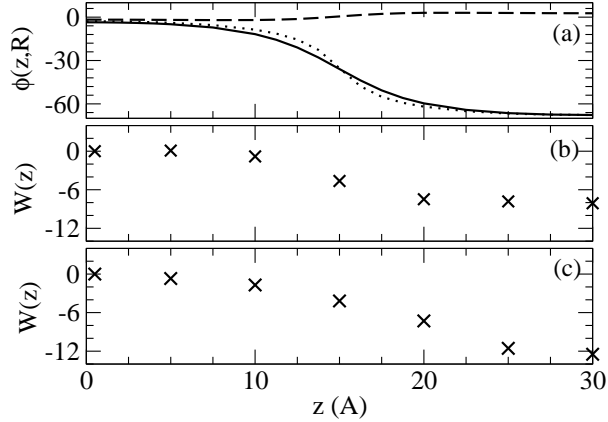}}}
\caption[]
{\label{fig2} \noindent
Na$^+$ potential of mean force $W(z)$~kcal/mol along the pore axis, with 
$\phi_{\rm d}|e|=71.7$~kcal/mol, $L=60$~\AA, and a unit cell doubled in
the $z$-direction.  (a) No water; solid, dotted, and dashed lines
refer to ions at radii $R/R_{\rm d}$=0, 0.7, and 1.8 respectively. (b)
Water-filled tube geometry. (c) Water-filled membrane geometry.}
\end{figure}

We next use two models to demonstrate how this intrinsic $\phi_d$
may be exploited: (1) an infinitely long ``tube'' geometry, with
repeating unit cells that have alternating $L/2$-long stretches where
dipoles surface densities exist or vanish (Fig.~\ref{fig1}a); (2) a ``membrane''
geometry with dipoles in the pore interior only (Fig.~\ref{fig1}b).
Regions with low local dipole density can potentially be achieved by
functionalizing silica pores with amine groups.\cite{pore}
Near the junction, $z=z_o$, between dipolar ($\phi=-4\pi \eta_d d$)
and non dipolar ($\phi=0$) regions, the pore-axis electrostatic potential is
\begin{equation}
\phi_{\rm d}(z,R=0)= -2\pi \eta_d d \bigg\{ \frac{(z-z_o)/R_{\rm d}}
                { \sqrt{1+[(z-z_o)/R_d]^2} }+1 \bigg\} \label{eq1}
\end{equation}
when $L$ far excceds $|z-z_o|$ and $R_{\rm d}$. 
Figure~\ref{fig2}a depicts $\phi(z,R)$ for $R_{\rm d}=5.5$~\AA, $L=60$~\AA.
$\phi(z,R)$ is fairly independent of $R$ inside the pore but vanishes
rapidly for $R > R_{\rm d}$.  An electric field {\it parallel} to the
pore interior surface materializes,\cite{field} localized within a distance
$R_{\rm d}$ of the entrance.  This field is locally screened by
water molecules.  

Note that in the membrane model,
if the membrane surface has the same dipole density as the pore interior
surface, it exerts an equal and opposite $\phi_{\rm d}(z,R=0)$ that
cancels Eq.~\ref{eq1}.\cite{support}  Hence, to exploit dipolar effects, the
two surfaces must be functionalized differently. 
 
To examine ion permeation into water-filled dipolar pores,
we apply the atomistic H$_2$O, Na$^+$, and Cl$^-$ force fields used in
Ref.~\onlinecite{pore} to compute the potential of mean force
$W(z)$.  Pore-water/ion interactions are described
by analytical, least square fit approximations to dipolar electrostatic
and Lennard-Jones terms.\cite{support}  The pores are hydrophilic, with
$\sim 12$~\AA\, diameter and 2.7~\AA$^{-1}$ H$_2$O density.
Infinitely long but laterally isolated tubes (Fig.~\ref{fig1}a)
are simulated using at least 10$^5$ passes of canonical ensemble Monte Carlo
(MC) runs.  Membrane models (Fig.~\ref{fig1}b), periodic in all 3 directions
with lateral unit cell lengths of 40~\AA, are simulated for at least
200~ps using molecular dynamics (MD)/particle mesh Ewald methods.
Simulation cells contain two unit cells doubled in the $z$ direction.
Because Na$^+$ orders water in nanpores in a way incompatible with the
periodic boundary conditions used,\cite{dellago} our simulation cells
contain two unit cells doubled in the $z$ direction for convergence.

Figure~\ref{fig2}b depicts $W(z)$ for a $\phi_{\rm d}=-72$~kcal/mol
and $L/2=30$~\AA\, tube.  The membrane geometry 
leads to qualitatively similar results (Fig.~\ref{fig2}c). These
geometries yield $\Delta W_1= W(L/2)- W(0)=-8.1$ and $-12.5$~kcal/mol.
Compared with $\Delta W_1=\phi_d$=-72~kcal/mol in the absence of water,
they exhibit effective dielectric constants of 8.9 and 6.0, respectively.
 
However, unlike transmembrane ion channels such as KcsA,\cite{kcsa}
material pores can have lengths on the order of 1 $\mu$m.  Table~\ref{table1}
shows that $\Delta W_1$ in the tube geometry steadily increases with
$L/2$ and is roughly proportional to $\phi_{\rm d}$; at $L/2=120$~\AA,
$\epsilon \sim 2.3$.  This is readily explained as follows.  Unless the
localized electric field induces a ferroelectric phase transition (not
the case here), $\Delta W_1$ is predominantly screened by H$_2$O
near the junctions where the electric field at its maximum.  For
large $L$, these localized H$_2$O resemble a point dipole whose
interaction with ions vanishes as $1/|z-z_o|^2$.  Thus, the attraction
of cations deep inside dipolar regions is asymptotically {\it unscreened}
by water.  $\Delta W_1$ analysis for the membrane geometry
is complicated by its non-zero value even when the dipoles are absent
in membrane slabs, and by the $O(1)$~kcal/mol statistical noise,
but $\Delta W_1$ follows a qualitatively similar trend there.
 
As long as $L \gg R_d$, the screening should asymptotically vanish
even when $R_d$ is large.  In macroscopic systems, it will
be the ever-present electrolytes in water, not water itself, that
screen the ion-surface dipole interaction.  Here nanoscale pores with
$R_d < 10$~\AA\, may exhibit unique properties.  (1) We consider 
1.0~M NaCl electrolytes in the membrane geometry.  The hydration
free energy of the Na$^+$ in the middle of the dipolar pore is reduced
(last lines of Table~\ref{table1}) because other Na$^+$ ions now enter
the nanopore, while Cl$^-$ seldom do; the resulting electrostatic potential
increase makes cations less favorable.\cite{support}  But unlike
macroscopic pores, even the 1.0~M NaCl electrolytes do not yield complete
screening in this membrane geometry.  This is possibly due to confinement
effects.
Results from 2.7 and 5.1~ns MD trajectories with
different initial configurations are used here because ions diffuse
rather slowly.  (2) Azobenzene\cite{blockers}
or similar functional groups can be placed at the junctions where the
axial electric field is largest, further restricting water density there and
hindering screening.  (3) Nanopores with alternating dipolar and non-dipolar
surfaces may result in an ionic, charge segregated, insulating behavior.

We use MC to demonstrate that such a ``phase separation'' occurs
with a confined 0.5~M NaCl inside an infinitely long pore with unit cell
$L=240$~\AA\, and $\phi_{\rm d} |e|=-72$~kcal/mol (see Fig.~\ref{fig1}c). 
The effective dipole-ion screening for this $L$ is 2.3 (Table~\ref{table1}).  
This Na$^+$ concentration is qualitatively consistent with our simulation
of permeation of 1.0~M NaCl into the nanopore in the membrane
geometry.\cite{support}
To accelerate ion sampling, water is here approximated as a dielectric
with $\epsilon_{\rm w}=10$, likely slightly underestimated
because this is obtained by considering a contact ion pair in the
tube.\cite{support} We find that ion transport is blocked even upon
applying a 0.0025 V/\AA\, electric field screened by $\epsilon_{\rm w}=10$.  
We stress that this insulating behavior arises
from (1) the almost-unscreened ion-pore dipole interaction dominating
the better-screened ($\epsilon_w=10$), ion-ion coulomb interactions, and
(2) the alternating dipolar regions leading to a complete absence of
cation/anion along lengths of the nanopore.  Future simulations of $I-V$
characteristics, taking into account leakage of Na$^+$ through the
nondipolar regions at long times, will be profitable.\cite{gillespie}
 
In conclusion, dipole layers naturally occur in certain nanopores, leading to
intrinsic preference for cations or anions.  This preference can be
exploited when the membrane surface is functionalized differently from
the pore interior, or when there are alternating dipolar/non-dipolar stretches
inside a long pore.  The dipole-ion interaction is asymptotically
unscreened by water, potentially leading to novel ionic, charge segregated,
insulating behavior that can block ion transport.  Our finding
may lead to new synthetic ion-exclusion membranes.
 
Sandia is a multiprogram laboratory operated by Sandia Corporation,
a Lockheed Martin Company, for the U.S.~Department of Energy's National
Nuclear Security Administration under contract DE-AC04-94AL8500.

\newpage 

\vspace*{2.5in}

\newpage 

\vspace{5.5in}
 
\newpage 

\begin{table}\centering
\begin{tabular}{  c c r c c c r  } \hline
& tube  & & & & mem. & \\ \hline
$\phi_{\rm d} |e|$ & $L/2$ & $\Delta W_1$ &  &
        $\phi_{\rm d} |e|$ &  $L/2$ & $\Delta W_1$ \\ \hline
-71.7 &  30  &  -8.1 & &-71.7 &  30 & -12.5 \\
-71.7 &  60  & -14.9 & &-71.7 &  60 & -17.4 \\
-71.7 &  90  & -24.0 & &  0.0 &  30 &  +3.3 \\
-71.7 & 120  & -32.3 & &  0.0 &  60 &  +2.5 \\ \hline
-23.9 &  60  &  -3.0 & &-23.9 &  30 &  -1.7 \\ 
-23.9 & 120  &  -8.1 & &  0.0 &  30 &  +3.3 \\ \hline
& & & $^*$NaCl &-71.7 &  30 &  -3.4  \\
& & & $^*$NaCl &  0.0  & 30 &  +2.6 \\ \hline
\end{tabular}
\caption[]
{\label{table1} \noindent
Na$^+$ potential of mean force ($\Delta W_1$) at $z=L/2$,
relative to its value at $z=0$ where $\phi_{\rm d}\approx 0$.
Energies and lengths are in units of kcal/mol and \AA.
$^*$Computed in 1.0~M NaCl electrolyte, not pure water.
}
\end{table}

\end{document}